\documentclass[%
 aip,
 amsmath,amssymb,
 reprint,%
]{revtex4-1}

\usepackage{graphicx}
\usepackage{dcolumn}
\usepackage{bm}

\usepackage[utf8]{inputenc}
\usepackage[T1]{fontenc}
\usepackage{mathptmx}
\usepackage{etoolbox}
\usepackage{comment}
\usepackage{xcolor}

\makeatletter
\def\@email#1#2{%
 \endgroup
 \patchcmd{\titleblock@produce}
  {\frontmatter@RRAPformat}
  {\frontmatter@RRAPformat{\produce@RRAP{*#1\href{mailto:#2}{#2}}}\frontmatter@RRAPformat}
  {}{}
}%
\makeatother
\begin{document}

\preprint{AIP/123-QED}

\title{Mid-infrared optical waveguides: missing link of integrated photonics}

\author{T Toney Fernandez$^{*}$}
    \altaffiliation{These authors contributed equally to this work.}
\author{Yongsop Hwang}
    \altaffiliation{These authors contributed equally to this work.}
\author{D E Otten}
    \email{toney.teddyfernandez@unisa.edu.au}
        \affiliation{ 
        STEM, University of South Australia, Australia
                    }

\author{A Fuerbach}
        \affiliation{ 
        MQ Photonics Research Centre, School of Mathematical and Physical Sciences, Macquarie University, NSW, 2109, Australia
                    }
\author{D G Lancaster}
        \affiliation{ 
        STEM, University of South Australia, Australia
                    }

\date{\today}

\begin{abstract}
A lack of low phonon energy glass waveguides delays the progress of integrated photonics in the mid-infrared. This perspective provides insights on the history of waveguide production in the past and what to expect in the near future.
\end{abstract}

\maketitle

\section{\label{sec:level1}Introduction}
\noindent The integrated photonic industry has long pursued transparent and amorphous waveguides operating in the mid-infrared wavelength range as a highly desired asset~\cite{Hu2019}. Until now, all proposed or anticipated techniques aimed at achieving these objectives have not succeeded. Consequently, there are no mid-infrared waveguide lasers or waveguide optical elements integrated into fiber architectures. Tellurites, chalcogenides and fluoride glasses are the suitable materials for mid-infrared wavelengths. Chalcogenides and tellurites were successful to some extent but the inherently high refractive index of these glasses produce high Fresnel/return losses and also induces detrimental non-linear effects~\cite{Hu2019}. Recently, there has been notable progress and revitalization in mid-infrared photonic technologies, particularly with the utilization of fluoride optical fibers. These advancements in fluoride fibers can be attributed to their exceptional broad transparency, low phonon energy, low bulk refractive index, and robust thermal and mechanical properties, surpassing those of other low phonon glasses. An exhaustive review and forecast of mid-infrared fiber laser research can be found in reference ~\cite{Jackson2024}. The typical phonon energy of a fluoride glass is between 400-600 cm$^{-1}$ (depending on the glass former), which is less than half of the phonon energy of silicates ($\approx1100$~cm$^{-1}$), phosphates ($\approx 1250$~cm$^{-1}$), or germanates ($\approx 900$~cm$^{-1}$), \textit{c.f} Table 4 in~\cite{Gao2017}. This de-thermalization of the rare-earth energy levels in fluoride glasses leads to unique and novel laser transitions, resulting in pioneering laser generation in wavelengths from the visible to mid-infrared~\cite{Bernier2022}. However, these results could not be fully exploited in optical waveguides because the state of the art production of waveguides in fluoride glasses are based on depressed claddings that guide a tunnelling mode with minimal confinement and low NA. This works well in the mid infrared region with a trade-off producing large laser output modes of 50-100~$\mu$m~\cite{Lancaster2011, Lancaster2012}. But when the operating wavelength is shifted to NIR and visible, there are no successful waveguides reported in fluoride glasses as depressed cladding waveguides that operates in visible wavelength were physically impossible to fabricate. A higher spatial confinement is required for novel visible rare-earth transitions as most of them have a self-terminating behaviour where the upper laser level has a lifetime equal or shorter than the lower laser level.  When analysing the low gain transitions it can be seen that forcing those transitions to lase will always end up in pulsed operation~\cite{Al-Mahrous, JACKSON2022}. 
The expansive range of light generation spanning from UV to mid-infrared provided by fluoride glass presents a remarkable color palette promising for displays, materials processing, biomedical applications, environmental sensing, optical communications. If passive optical elements like splitters, combiners, gratings, modulators, polarizers, resonators, switches etc can also be designed in the same glass, then a fully integrated system could be designed and novel devices could be realized.

\section{\label{sec:level1}First generation fluoride glass waveguides}
\begin{figure}[t]
    \centering
    \includegraphics[trim={0 0 0 0},width=5 cm]{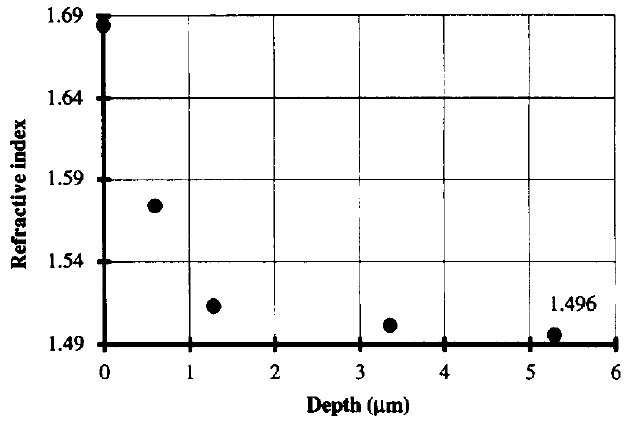}
        \caption{Index profile obtained in a fluoride glass through Na$^{+}$ : Li$^{+}$ ion diffusion. Reproduced with permission~\cite{FOGRET199679}. }
    \label{Index_Profile}
    \end{figure}

In the 1990s, anion and cation exchanges were employed to fabricate waveguides in fluoride glass. Various ions such as OH$^{-}$, OD$^{-}$ (where D represents Deuterium), Cl$^{-}$, Li$^{+}$, Na$^{+}$, and K$^{+}$ were investigated (table~\ref{table:1}). Among them, chlorine showed the most promising results, with reported positive index changes as high as $1 \times 10^{-1}$. Despite a linear decrease in density as fluoride was replaced by chloride, chloride's significantly higher polarizability, exceeding 350\%, enabled this success (Figure~\ref{Index_Profile}).

However, cationic exchange conducted in fluoride glass using alkali-containing inorganic molten salts led to severe devitrification and rapid corrosion of the samples. Diluted nitrates in organic solvents helped mitigate degradation to some extent but failed to produce waveguiding structures. ZBLAN (ZrF$_4$-BaF$_2$) and BIG (BaF$_2$-InF$_3$) served as the host platforms for these experiments. Among monovalent cations, only Li$^+$, Na$^+$, and K$^+$ were identified as potential candidates for incorporation into BIG fluoride glasses. As the diffusion process needed to maintain the exchanged layer vitreous, only Li$^+$ and K$^+$ could serve as suitable incoming cations.\par
When K$^+$ replaced Na$^+$, the predominant effects included the generation of compressive stresses and the introduction of ions with higher electronic polarizability. Conversely, when Li$^{+}$ replaced Na$^{+}$, the primary effect was an increase in the electronic density of the exchanged layer. Consequently, both of these processes proved suitable for creating waveguiding structures.

\begin{table}[h!]
\centering
\begin{tabular}{c c c c c c} 
 \hline
 Ref. & Glass & Ion & Exchange time & $\Delta n$ & Depth\\ 
 & & Out:In & (hours)& &($\mu$m)\\
 \hline
 ~\cite{FOGRET199679} & BIG & Na$^{+}$ : Li$^{+}$ & 16 & $2 \times 10^{-1}$ & 5.3\\
~\cite{JOSSE19971139} & ZBLA & ~~F$^{-}$ : Cl$^{-}$ & 10.5 & $1 \times 10^{-1}$ & 9.5\\ 
 ~\cite{JOSSE1997152} & BIG & ~~~F$^{-}$ : OD$^{-}$ & 25 & $4.8 \times 10^{-2}$ & 6\\ 
 ~\cite{JOSSE1997152} & ZBLA & ~F$^{-}$ : Cl$^{-}$ & 72 & $2.7 \times 10^{-2}$ & 14\\
 ~\cite{SRAMEK1999189} & ZBLA & ~F$^{-}$ : Cl$^{-}$ & 10.5 & $3 \times 10^{-2}$ & 10\\ 
 \hline
\end{tabular}
\caption{Reported index changes in fluoride glasses by ion exchange technique}
\label{table:1}
\end{table}

Most of the reports were planar waveguides and the major issue in producing channel waveguides in these heavy metal fluoride glasses were the lack of control and adhesion of masking material to the glass. For the barium-indium-gallium glass~\cite{JOSSE1997152}, aluminium and a diffusion mask was successfully deposited using photolithography but aluminium metal reduced the trivalent indium component within the bulk glass ending up corroding the surface making it impossible for ion-exchange. Zn, Cr, Si and Au were tried and only silica mask was compatible for BIG glass. Aluminium mask was used for zirconium-barium-lanthanum-aluminium glass.

\section{\label{sec:level1}State-of-the-art fluoride glass waveguides}
\begin{figure}[t]
    \centering
    \includegraphics[trim={0 0 0 0},width=6 cm]{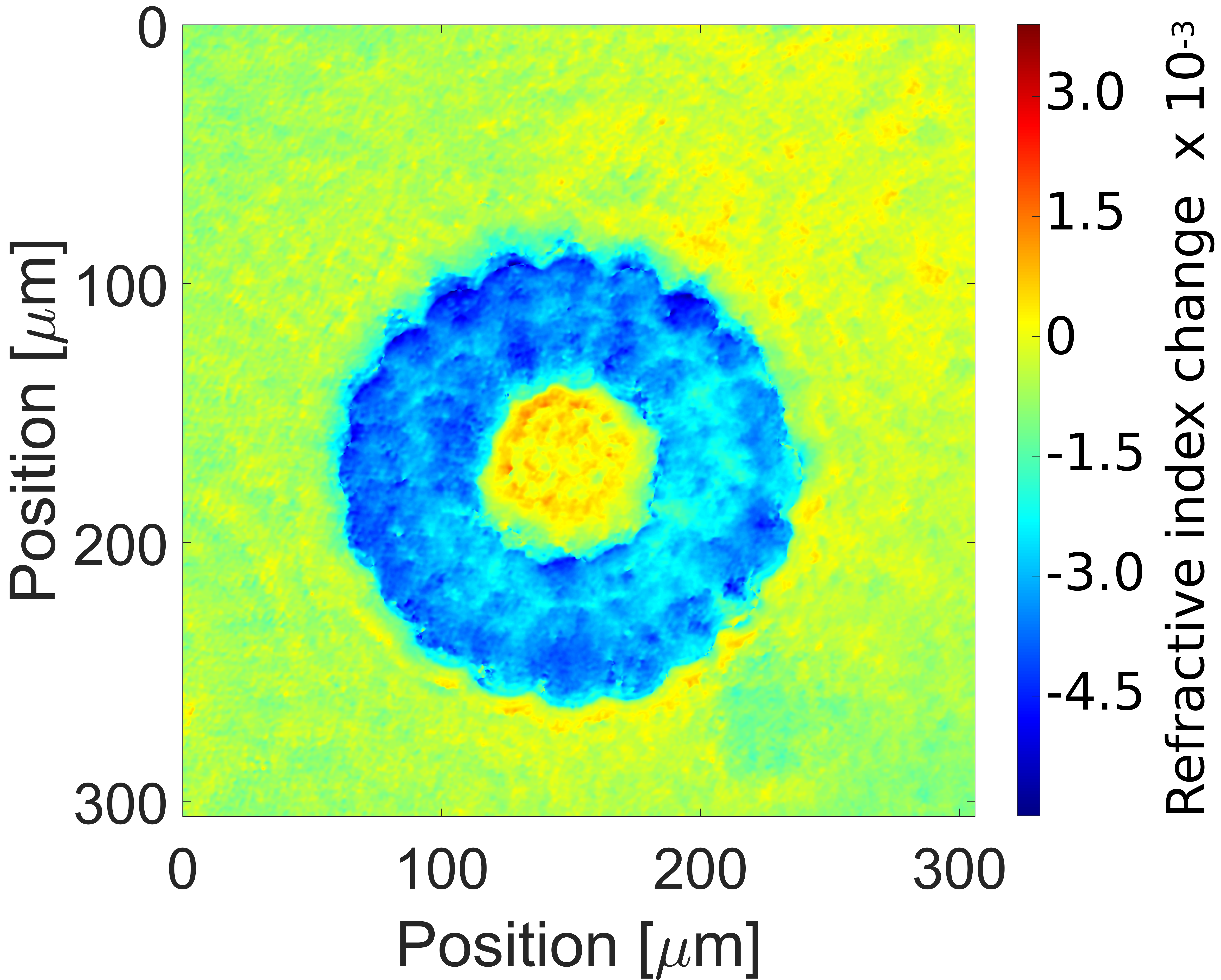}
        \caption{Index profile of a depressed cladding waveguide fabricated via ultrafast laser micromachining.}
    \label{Phase}
    \end{figure}

Masking remains a significant challenge for ion-exchanged channel waveguides in heavy metal glasses, prompting a shift towards maskless techniques. Ultrafast laser micromachining has emerged as the primary area of research for inscribing waveguides in these low phonon glasses. A pioneering study, led by K. Miura et al.~\cite{MIURA1999212}, demonstrated the tunability of waveguide width and index change by adjusting pulse widths and conducting multiple scan passes. They reported a positive index change of $3 \times 10^{-3}$. However, subsequent reports were scarce, primarily due to the fact that mid-infrared wavelengths require high refractive index changes and/or large areas of modification to meet the necessary V-number for channeled guiding ~\cite{snyder1983optical}. Attempts to fabricate larger waveguides resulted in ultrashort pulses interacting with fluoride glass, leading to the creation of negative index change zones. To take advantage of this unfortunate result, a structurally engineered approach known as the depressed cladding structure was developed whereby concentric negative index changes are written around an unmodified central guiding zone. There are several drawbacks in designing devices fabricated with such waveguides 1) The resulting larger mode area necessitates a higher pump power to achieve complete inversion within the waveguide for lasing, which directly impacts the optical-to-optical slope efficiency 2) controlling or manipulating large output modes poses challenges such as reduced divergence, low spatial coherence, non-uniform intensity profiles, and unmatched pump to signal modes for lasing 3) low refractive index change attributed to depressed cladding waveguides makes it difficult to inscribe devices that requires larger bend radii, the smallest bend radius reported for such a waveguides is around 313 mm~\cite{Gross2015} 4)creating waveguides operable at long wavelengths involves inscribing large structures that must be buried deep within the glass substrate, they are potentially susceptible to strong nonlinear and aberration effects. Additionally, objectives with long working distances tend to be expensive. 5) high coupling loss when coupled to any commercial optical fibers. The refractive index map of a typical depressed cladding waveguide written in ZBLAN glass using a high repetition rate laser is illustrated in Figure~\ref{Phase}.
Other efforts in producing larger V-numbers suitable for mid infrared wavelengths fell short with either low index changes and/or smaller guiding zones. The consolidated results of such previous reports are shown in table~\ref{table:2}

\begin{table}[h!]
\centering
\begin{tabular}{c c c c c c} 
 \hline
 Ref. & Glass & Rep.Rate & Feed rate & $\Delta$n & Width\\ 
 & & kHz & mm/s & &($\mu$m)\\
 \hline
 ~\cite{Berube13} & ZHBLAN & 1-250 & 0.05-5 & $1.25 \times 10^{-3}$ & 10\\
 ~\cite{Ledemi14} & Fluoroborate & 100 & 0.05-50 & $5 \times 10^{-3}$ & irregular\\ 
 ~\cite{Fernandez2022} & ZHBLAN & 5-50 & 0.05-0.2 & $6.5 \times 10^{-3}$ & 12-20\\ 
 \hline
\end{tabular}
\caption{Reported index changes in fluoride glasses by ion exchange technique}
\label{table:2}
\end{table}

\section{\label{sec:level1} Ideal mid-infrared waveguide: design expectations}

\noindent It is desirable to have a channel waveguide 
with a high-index core in a low-index background bulk material, 
commonly called a type 1 waveguide,
instead of a depressed cladding for multiple reasons.
A depressed cladding waveguide in a fluoride glass tends to
have a large core diameter due to low index contrast and/or difficulties in arranging large negative index changes around a small core.
A high index contrast is also essential 
to reduce the guided mode diameter
so that the waveguidde can be integrated
with other photonic devices.
A depressed cladding waveguide also has geometric limitation
which makes additional functionalities
such as integrating with a ring resonator
or forming a Mach-Zehnder interferometer
by bringing two bent waveguides close to each other.
A type 1 waveguide has benefits in terms of the geometric flexibility, including both the ability to easily bring two guiding regions into close proximity and....

\begin{figure}[htp!]
    \includegraphics[width=0.85\columnwidth]{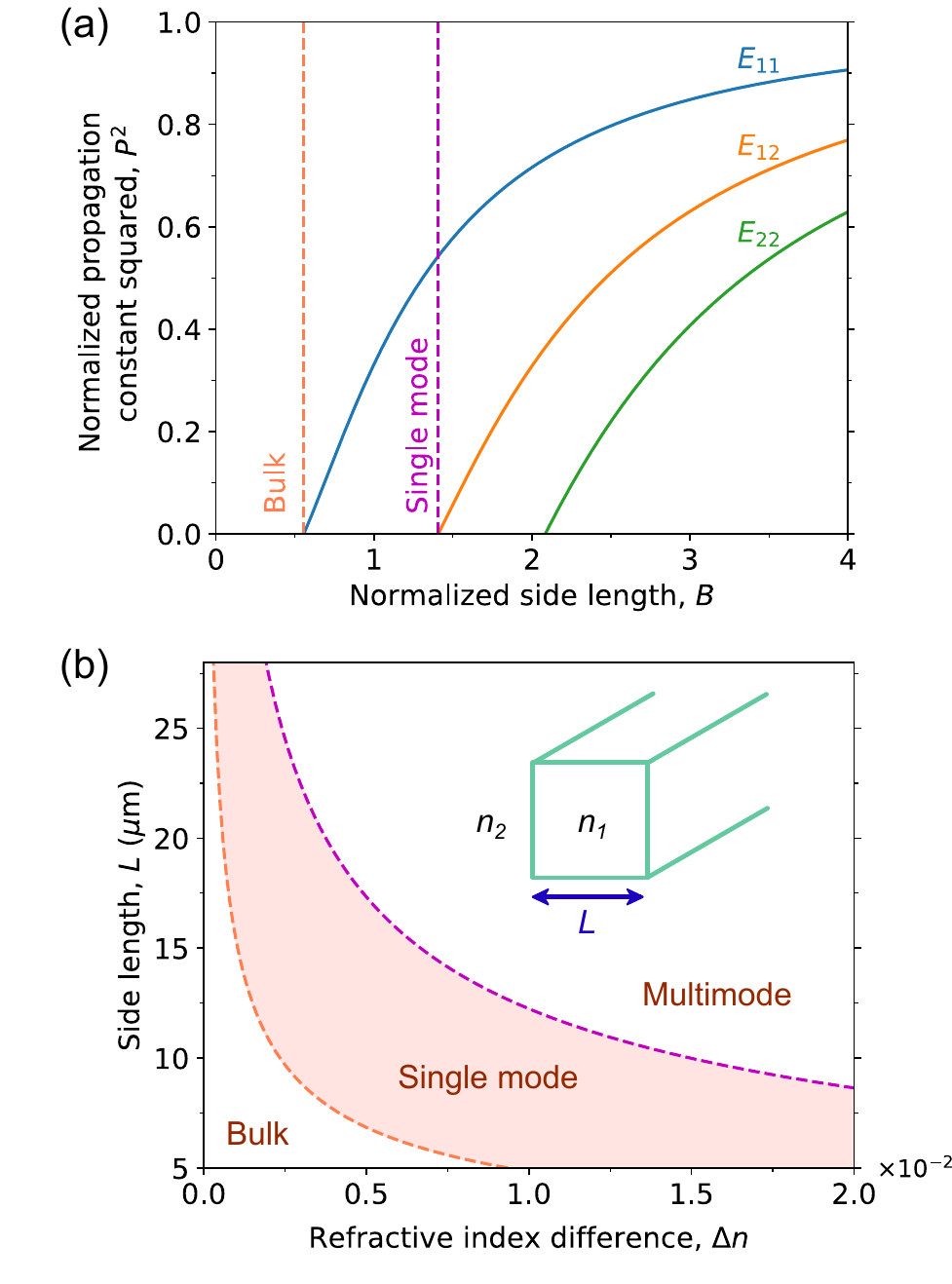}
    \caption{(a) The dispersion curves of the three lowest guided modes: 
    $E_{11}$, $E_{12} (=E_{21}$), and $E_{22}$.
    The cutoffs for single mode and bulk
    are indicated by the vertical dashed lines.
    (b) The design condition to satisfy 
    for a single-mode waveguide
    is indicated for a square waveguide of ZBLAN
    at the wavelength of 3~$\mu$m.
    The index difference $\Delta n = n_1 - n_2$.
    }
    \label{fig:single}
\end{figure}

The required dimensions of such waveguides can be 
estimated theoretically.
Here we present our estimation of a mid-infrared waveguide
as shown in Fig.~\ref{fig:single} based on 
the theory of rectangular waveguides.~\cite{kumar1983analysis}
Assume that we have a square waveguide with a side length of $L$
with a core and a cladding indices of $n_1$ and $n_2$, respectively,
as shown in the inset of Fig.~\ref{fig:single}(b).
When $n_1 > n_2$, the dispersion relation 
of the supported modes can be obtained 
using a perturbation approach introduced by Kumar \textit{et al.}~\cite{kumar1983analysis}
We identify the guided modes following the $E_{pq}$ convention
where $p$ and $q$ are horizontal and vertical mode numbers, respectivly.

The parameter shown in the $x$-axis is
the normalized side length
which is defined as follows:
\begin{equation}
    B=\frac{k_0 L}{\pi} (n_1 - n_2)^{1/2},
    \label{eq:B}
\end{equation}
where $k_0$ and $L$ are the free space wavenumber 
and the side length of the square waveguide, respectively.
The propagation constant is also advantageous to be 
defined in as a dimensionless quantity for generality.
\begin{equation}
    P^2 = P_0^2 + P^{\prime2}.
\end{equation}
Here, $P_0$ and $P^\prime$ are the unperturbed normlized propagation constant
and the perturbation term, respectively,
which are defined as 
\begin{eqnarray}
    P_0^2 &=& \frac{\beta^2-k_0^2 n_2 ^2}{k_0^2 (n_1^2-n_2^2)},\qquad\text{and} \\
    P^{\prime2} &=& \frac{1}{n_1^2-n_2^2}
        \left[\frac{4\int_{L/2}^\infty\int_{L/2}^\infty|\psi|^2(n_1^2-n_2^2)dx dy}
        {\int_{-\infty}^\infty\int_{-\infty}^\infty |\psi|^2 dx dy} \right].
\end{eqnarray}

The dispersion curves of the lowest three guided modes:
$E_{11}$, $E_{12}(=E_{21})$, and $E_{22}$ 
are plotted in Fig.~\ref{fig:single}(a).
The cutoff for single mode is indicated
which implies that when the normalized side length is larger than 
the cutoff higher order modes can resize in the waveguide.
The cutoff for bulk is also indicated,
if the normalized side length is smaller than the cutoff,
then there will be no waveguide mode.

Using the cutoff values in the dispersion curves and Eq.~(\ref{eq:B}),
the desired range of the side length 
and the refractive index difference, $\Delta n = n_1 - n_2$,
can be obtained for a single-mode waveguide as shown in Fig.~\ref{fig:single}(b).
The single-mode region of a ZBLAN (53\% ZrF$_4$) waveguide at 3~$\mu$m is shaded,
when the cladding index $n_2=1.4817$.~\cite{gan1995optical}
We can notice that
the waveguide size must be incresed to support a guided mode
when the index contrast is low, 
since otherwise, only the bulk guiding will be possible.
Considering a scenario of integration with a single-mode fiber
with a core diamater in the range of 5-8~$\mu$m as commonly available,
a refactive index difference $\Delta n > 0.01$ is generally desired. 

\section{Conclusion}
In conclusion, the development of fluoride glass waveguides for mid-infrared photonic applications represents a critical advancement in integrated photonics. Historically, challenges related to the production and optimization of these waveguides—such as low refractive index contrast and difficulties in fabrication—have impeded progress. However, recent innovations, particularly in ultrafast laser micromachining, offer promising pathways to overcome these barriers.

For the future, the focus should be on refining waveguide designs to achieve higher refractive index contrasts and better mode confinement, which are crucial for enabling the integration of these waveguides into broader photonic systems. The realization of type 1 waveguides, with high-index cores in low-index background materials, is particularly promising for achieving the necessary confinement and scalability in integrated photonic circuits. This would not only enhance the performance of mid-infrared devices but also expand their applicability across a range of fields, including telecommunications, biomedical sensing, and environmental monitoring.

The theoretical framework presented in this manuscript outlines the parameters necessary for designing efficient single-mode mid-infrared waveguides, providing a solid foundation for future experimental work. By achieving higher index changes and optimizing waveguide dimensions, it is possible to create more effective photonic devices, paving the way for significant advancements in mid-infrared photonics and integrated optical systems.

\begin{acknowledgments}
DL acknowledges the support by the Australian Research Council COMBS Centre of Excellence (CE230100006). TTF, YH and DL acknowledge support received from Electro Optic Systems Pty. Limited. Authors gratefully acknowledge the access of Digital Holographic Microscope from South Australian Node of the Australian National Fabrication Facility at Mawson Lakes.

\end{acknowledgments}
This research was supported by the Australian Research Council Centre of Excellence in Optical Microcombs for Breakthrough Science (project number CE230100006) and funded by the Australian Government. TTF, YH and DL acknowledge support received from Electro Optic Systems Pty. Limited. Authors acknowledge the use of Digital Holographic Microscope facility at  Australian National Fabrication Facility, SA Node, Mawson Lakes.

\section*{Data Availability Statement}

The data that support the findings of this study are available from the corresponding author upon reasonable request.
\\
\\
\section*{References}

\bibliography{References}

\begin{thebibliography}{20}%
\makeatletter
\providecommand \@ifxundefined [1]{%
 \@ifx{#1\undefined}
}%
\providecommand \@ifnum [1]{%
 \ifnum #1\expandafter \@firstoftwo
 \else \expandafter \@secondoftwo
 \fi
}%
\providecommand \@ifx [1]{%
 \ifx #1\expandafter \@firstoftwo
 \else \expandafter \@secondoftwo
 \fi
}%
\providecommand \natexlab [1]{#1}%
\providecommand \enquote  [1]{``#1''}%
\providecommand \bibnamefont  [1]{#1}%
\providecommand \bibfnamefont [1]{#1}%
\providecommand \citenamefont [1]{#1}%
\providecommand \href@noop [0]{\@secondoftwo}%
\providecommand \href [0]{\begingroup \@sanitize@url \@href}%
\providecommand \@href[1]{\@@startlink{#1}\@@href}%
\providecommand \@@href[1]{\endgroup#1\@@endlink}%
\providecommand \@sanitize@url [0]{\catcode `\\12\catcode `\$12\catcode `\&12\catcode `\#12\catcode `\^12\catcode `\_12\catcode `\%12\relax}%
\providecommand \@@startlink[1]{}%
\providecommand \@@endlink[0]{}%
\providecommand \url  [0]{\begingroup\@sanitize@url \@url }%
\providecommand \@url [1]{\endgroup\@href {#1}{\urlprefix }}%
\providecommand \urlprefix  [0]{URL }%
\providecommand \Eprint [0]{\href }%
\providecommand \doibase [0]{http://dx.doi.org/}%
\providecommand \selectlanguage [0]{\@gobble}%
\providecommand \bibinfo  [0]{\@secondoftwo}%
\providecommand \bibfield  [0]{\@secondoftwo}%
\providecommand \translation [1]{[#1]}%
\providecommand \BibitemOpen [0]{}%
\providecommand \bibitemStop [0]{}%
\providecommand \bibitemNoStop [0]{.\EOS\space}%
\providecommand \EOS [0]{\spacefactor3000\relax}%
\providecommand \BibitemShut  [1]{\csname bibitem#1\endcsname}%
\let\auto@bib@innerbib\@empty
\bibitem [{\citenamefont {Hu}\ and\ \citenamefont {Yang}(2019)}]{Hu2019}%
  \BibitemOpen
  \bibfield  {author} {\bibinfo {author} {\bibfnamefont {J.}~\bibnamefont {Hu}}\ and\ \bibinfo {author} {\bibfnamefont {L.}~\bibnamefont {Yang}},\ }\enquote {\bibinfo {title} {Glass in integrated photonics},}\ in\ \href {\doibase 10.1007/978-3-319-93728-1_42} {\emph {\bibinfo {booktitle} {Springer Handbook of Glass}}},\ \bibinfo {editor} {edited by\ \bibinfo {editor} {\bibfnamefont {J.~D.}\ \bibnamefont {Musgraves}}, \bibinfo {editor} {\bibfnamefont {J.}~\bibnamefont {Hu}}, \ and\ \bibinfo {editor} {\bibfnamefont {L.}~\bibnamefont {Calvez}}}\ (\bibinfo  {publisher} {Springer International Publishing},\ \bibinfo {address} {Cham},\ \bibinfo {year} {2019})\ pp.\ \bibinfo {pages} {1441--1481}\BibitemShut {NoStop}%
\bibitem [{\citenamefont {Jackson}(2024)}]{Jackson2024}%
  \BibitemOpen
  \bibfield  {author} {\bibinfo {author} {\bibfnamefont {S.~D.}\ \bibnamefont {Jackson}},\ }\bibfield  {title} {\enquote {\bibinfo {title} {{Mid-infrared fiber laser research: Tasks completed and the tasks ahead}},}\ }\href {\doibase 10.1063/5.0220406} {\bibfield  {journal} {\bibinfo  {journal} {APL Photonics}\ }\textbf {\bibinfo {volume} {9}},\ \bibinfo {pages} {070904} (\bibinfo {year} {2024})},\ \Eprint {http://arxiv.org/abs/https://pubs.aip.org/aip/app/article-pdf/doi/10.1063/5.0220406/20078771/070904\_1\_5.0220406.pdf} {https://pubs.aip.org/aip/app/article-pdf/doi/10.1063/5.0220406/20078771/070904\_1\_5.0220406.pdf} \BibitemShut {NoStop}%
\bibitem [{\citenamefont {Gao}\ \emph {et~al.}(2017)\citenamefont {Gao}, \citenamefont {Turshatov}, \citenamefont {Howard}, \citenamefont {Busko}, \citenamefont {Joseph}, \citenamefont {Hudry},\ and\ \citenamefont {Richards}}]{Gao2017}%
  \BibitemOpen
  \bibfield  {author} {\bibinfo {author} {\bibfnamefont {G.}~\bibnamefont {Gao}}, \bibinfo {author} {\bibfnamefont {A.}~\bibnamefont {Turshatov}}, \bibinfo {author} {\bibfnamefont {I.~A.}\ \bibnamefont {Howard}}, \bibinfo {author} {\bibfnamefont {D.}~\bibnamefont {Busko}}, \bibinfo {author} {\bibfnamefont {R.}~\bibnamefont {Joseph}}, \bibinfo {author} {\bibfnamefont {D.}~\bibnamefont {Hudry}}, \ and\ \bibinfo {author} {\bibfnamefont {B.~S.}\ \bibnamefont {Richards}},\ }\bibfield  {title} {\enquote {\bibinfo {title} {Up-conversion fluorescent labels for plastic recycling: A review},}\ }\href {\doibase https://doi.org/10.1002/adsu.201600033} {\bibfield  {journal} {\bibinfo  {journal} {Advanced Sustainable Systems}\ }\textbf {\bibinfo {volume} {1}},\ \bibinfo {pages} {1600033} (\bibinfo {year} {2017})},\ \Eprint {http://arxiv.org/abs/https://onlinelibrary.wiley.com/doi/pdf/10.1002/adsu.201600033} {https://onlinelibrary.wiley.com/doi/pdf/10.1002/adsu.201600033} \BibitemShut {NoStop}%
\bibitem [{\citenamefont {Bernier}\ \emph {et~al.}(2022)\citenamefont {Bernier}, \citenamefont {Fortin}, \citenamefont {Henderson-Sapir}, \citenamefont {Jackson}, \citenamefont {Jobin}, \citenamefont {Li}, \citenamefont {Luo}, \citenamefont {Maes}, \citenamefont {Majewski}, \citenamefont {Ottaway}, \citenamefont {Vallée},\ and\ \citenamefont {Aydin}}]{Bernier2022}%
  \BibitemOpen
  \bibfield  {author} {\bibinfo {author} {\bibfnamefont {M.}~\bibnamefont {Bernier}}, \bibinfo {author} {\bibfnamefont {V.}~\bibnamefont {Fortin}}, \bibinfo {author} {\bibfnamefont {O.}~\bibnamefont {Henderson-Sapir}}, \bibinfo {author} {\bibfnamefont {S.}~\bibnamefont {Jackson}}, \bibinfo {author} {\bibfnamefont {F.}~\bibnamefont {Jobin}}, \bibinfo {author} {\bibfnamefont {J.}~\bibnamefont {Li}}, \bibinfo {author} {\bibfnamefont {H.}~\bibnamefont {Luo}}, \bibinfo {author} {\bibfnamefont {F.}~\bibnamefont {Maes}}, \bibinfo {author} {\bibfnamefont {M.~R.}\ \bibnamefont {Majewski}}, \bibinfo {author} {\bibfnamefont {D.~J.}\ \bibnamefont {Ottaway}}, \bibinfo {author} {\bibfnamefont {R.}~\bibnamefont {Vallée}}, \ and\ \bibinfo {author} {\bibfnamefont {Y.~O.}\ \bibnamefont {Aydin}},\ }\bibfield  {title} {\enquote {\bibinfo {title} {Chapter 9 - high-power continuous wave mid-infrared fluoride glass fiber lasers},}\ }in\ \href {\doibase https://doi.org/10.1016/B978-0-12-818017-4.00008-2} {\emph {\bibinfo
  {booktitle} {Mid-Infrared Fiber Photonics}}},\ \bibinfo {series and number} {Woodhead Publishing Series in Electronic and Optical Materials},\ \bibinfo {editor} {edited by\ \bibinfo {editor} {\bibfnamefont {S.}~\bibnamefont {Jackson}}, \bibinfo {editor} {\bibfnamefont {M.}~\bibnamefont {Bernier}}, \ and\ \bibinfo {editor} {\bibfnamefont {R.}~\bibnamefont {Vallée}}}\ (\bibinfo  {publisher} {Woodhead Publishing},\ \bibinfo {year} {2022})\ pp.\ \bibinfo {pages} {505--595}\BibitemShut {NoStop}%
\bibitem [{\citenamefont {Lancaster}\ \emph {et~al.}(2011)\citenamefont {Lancaster}, \citenamefont {Gross}, \citenamefont {Ebendorff-Heidepriem}, \citenamefont {Kuan}, \citenamefont {Monro}, \citenamefont {Ams}, \citenamefont {Fuerbach},\ and\ \citenamefont {Withford}}]{Lancaster2011}%
  \BibitemOpen
  \bibfield  {author} {\bibinfo {author} {\bibfnamefont {D.~G.}\ \bibnamefont {Lancaster}}, \bibinfo {author} {\bibfnamefont {S.}~\bibnamefont {Gross}}, \bibinfo {author} {\bibfnamefont {H.}~\bibnamefont {Ebendorff-Heidepriem}}, \bibinfo {author} {\bibfnamefont {K.}~\bibnamefont {Kuan}}, \bibinfo {author} {\bibfnamefont {T.~M.}\ \bibnamefont {Monro}}, \bibinfo {author} {\bibfnamefont {M.}~\bibnamefont {Ams}}, \bibinfo {author} {\bibfnamefont {A.}~\bibnamefont {Fuerbach}}, \ and\ \bibinfo {author} {\bibfnamefont {M.~J.}\ \bibnamefont {Withford}},\ }\bibfield  {title} {\enquote {\bibinfo {title} {Fifty percent internal slope efficiency femtosecond direct-written tm3$+$:zblan waveguide laser},}\ }\href {\doibase 10.1364/OL.36.001587} {\bibfield  {journal} {\bibinfo  {journal} {Opt. Lett.}\ }\textbf {\bibinfo {volume} {36}},\ \bibinfo {pages} {1587--1589} (\bibinfo {year} {2011})}\BibitemShut {NoStop}%
\bibitem [{\citenamefont {Lancaster}\ \emph {et~al.}(2012)\citenamefont {Lancaster}, \citenamefont {Gross}, \citenamefont {Fuerbach}, \citenamefont {Heidepriem}, \citenamefont {Monro},\ and\ \citenamefont {Withford}}]{Lancaster2012}%
  \BibitemOpen
  \bibfield  {author} {\bibinfo {author} {\bibfnamefont {D.~G.}\ \bibnamefont {Lancaster}}, \bibinfo {author} {\bibfnamefont {S.}~\bibnamefont {Gross}}, \bibinfo {author} {\bibfnamefont {A.}~\bibnamefont {Fuerbach}}, \bibinfo {author} {\bibfnamefont {H.~E.}\ \bibnamefont {Heidepriem}}, \bibinfo {author} {\bibfnamefont {T.~M.}\ \bibnamefont {Monro}}, \ and\ \bibinfo {author} {\bibfnamefont {M.~J.}\ \bibnamefont {Withford}},\ }\bibfield  {title} {\enquote {\bibinfo {title} {Versatile large-mode-area femtosecond laser-written tm:zblan glass chip lasers},}\ }\href {\doibase 10.1364/OE.20.027503} {\bibfield  {journal} {\bibinfo  {journal} {Opt. Express}\ }\textbf {\bibinfo {volume} {20}},\ \bibinfo {pages} {27503--27509} (\bibinfo {year} {2012})}\BibitemShut {NoStop}%
\bibitem [{\citenamefont {Al-Mahrous}(2012)}]{Al-Mahrous}%
  \BibitemOpen
  \bibfield  {author} {\bibinfo {author} {\bibfnamefont {R.}~\bibnamefont {Al-Mahrous}},\ }\enquote {\bibinfo {title} {All-fiber fluoride fiber lasers},}\ \ (\bibinfo  {publisher} {Cuvillier Verlag},\ \bibinfo {address} {Göttingen, Germany},\ \bibinfo {year} {2012})\ Chap.\ \bibinfo {chapter} {Chapter 1: Structure and Properties of Silica and Fluoride Glass Fiber}, pp.\ \bibinfo {pages} {5--16}\BibitemShut {NoStop}%
\bibitem [{\citenamefont {Jackson}\ \emph {et~al.}(2022)\citenamefont {Jackson}, \citenamefont {Tokita}, \citenamefont {{R. Majewski}}, \citenamefont {Henderson-Sapir}, \citenamefont {Ottaway}, \citenamefont {Vallée}, \citenamefont {Bernier},\ and\ \citenamefont {Maes}}]{JACKSON2022}%
  \BibitemOpen
  \bibfield  {author} {\bibinfo {author} {\bibfnamefont {S.}~\bibnamefont {Jackson}}, \bibinfo {author} {\bibfnamefont {S.}~\bibnamefont {Tokita}}, \bibinfo {author} {\bibfnamefont {M.}~\bibnamefont {{R. Majewski}}}, \bibinfo {author} {\bibfnamefont {O.}~\bibnamefont {Henderson-Sapir}}, \bibinfo {author} {\bibfnamefont {D.~J.}\ \bibnamefont {Ottaway}}, \bibinfo {author} {\bibfnamefont {R.}~\bibnamefont {Vallée}}, \bibinfo {author} {\bibfnamefont {M.}~\bibnamefont {Bernier}}, \ and\ \bibinfo {author} {\bibfnamefont {F.}~\bibnamefont {Maes}},\ }\bibfield  {title} {\enquote {\bibinfo {title} {Chapter 7 - spectroscopy of the rare-earth-ion transitions in fluoride glasses},}\ }in\ \href {\doibase https://doi.org/10.1016/B978-0-12-818017-4.00011-2} {\emph {\bibinfo {booktitle} {Mid-Infrared Fiber Photonics}}},\ \bibinfo {series and number} {Woodhead Publishing Series in Electronic and Optical Materials},\ \bibinfo {editor} {edited by\ \bibinfo {editor} {\bibfnamefont {S.}~\bibnamefont {Jackson}}, \bibinfo {editor}
  {\bibfnamefont {M.}~\bibnamefont {Bernier}}, \ and\ \bibinfo {editor} {\bibfnamefont {R.}~\bibnamefont {Vallée}}}\ (\bibinfo  {publisher} {Woodhead Publishing},\ \bibinfo {year} {2022})\ pp.\ \bibinfo {pages} {333--399}\BibitemShut {NoStop}%
\bibitem [{\citenamefont {Fogret}\ \emph {et~al.}(1996)\citenamefont {Fogret}, \citenamefont {Fonteneau}, \citenamefont {Lucas},\ and\ \citenamefont {Rimet}}]{FOGRET199679}%
  \BibitemOpen
  \bibfield  {author} {\bibinfo {author} {\bibfnamefont {E.}~\bibnamefont {Fogret}}, \bibinfo {author} {\bibfnamefont {G.}~\bibnamefont {Fonteneau}}, \bibinfo {author} {\bibfnamefont {J.}~\bibnamefont {Lucas}}, \ and\ \bibinfo {author} {\bibfnamefont {R.}~\bibnamefont {Rimet}},\ }\bibfield  {title} {\enquote {\bibinfo {title} {Fluoride glass planar optical waveguides by cationic exchange},}\ }\href {\doibase https://doi.org/10.1016/0925-3467(95)00057-7} {\bibfield  {journal} {\bibinfo  {journal} {Optical Materials}\ }\textbf {\bibinfo {volume} {5}},\ \bibinfo {pages} {79--86} (\bibinfo {year} {1996})}\BibitemShut {NoStop}%
\bibitem [{\citenamefont {Josse}, \citenamefont {Fonteneau},\ and\ \citenamefont {Lucas}(1997)}]{JOSSE19971139}%
  \BibitemOpen
  \bibfield  {author} {\bibinfo {author} {\bibfnamefont {E.}~\bibnamefont {Josse}}, \bibinfo {author} {\bibfnamefont {G.}~\bibnamefont {Fonteneau}}, \ and\ \bibinfo {author} {\bibfnamefont {J.}~\bibnamefont {Lucas}},\ }\bibfield  {title} {\enquote {\bibinfo {title} {Low-phonon waveguides made by {F-Cl-} exchange on fluoride glasses},}\ }\href {\doibase https://doi.org/10.1016/S0025-5408(97)00087-1} {\bibfield  {journal} {\bibinfo  {journal} {Materials Research Bulletin}\ }\textbf {\bibinfo {volume} {32}},\ \bibinfo {pages} {1139--1146} (\bibinfo {year} {1997})}\BibitemShut {NoStop}%
\bibitem [{\citenamefont {Josse}\ \emph {et~al.}(1997)\citenamefont {Josse}, \citenamefont {Broquin}, \citenamefont {Fonteneau}, \citenamefont {Rimet},\ and\ \citenamefont {Lucas}}]{JOSSE1997152}%
  \BibitemOpen
  \bibfield  {author} {\bibinfo {author} {\bibfnamefont {E.}~\bibnamefont {Josse}}, \bibinfo {author} {\bibfnamefont {J.}~\bibnamefont {Broquin}}, \bibinfo {author} {\bibfnamefont {G.}~\bibnamefont {Fonteneau}}, \bibinfo {author} {\bibfnamefont {R.}~\bibnamefont {Rimet}}, \ and\ \bibinfo {author} {\bibfnamefont {J.}~\bibnamefont {Lucas}},\ }\bibfield  {title} {\enquote {\bibinfo {title} {Planar and channel waveguides on fluoride glasses},}\ }\href {\doibase https://doi.org/10.1016/S0022-3093(97)00008-2} {\bibfield  {journal} {\bibinfo  {journal} {Journal of Non-Crystalline Solids}\ }\textbf {\bibinfo {volume} {213-214}},\ \bibinfo {pages} {152--157} (\bibinfo {year} {1997})}\BibitemShut {NoStop}%
\bibitem [{\citenamefont {Sramek}\ \emph {et~al.}(1999)\citenamefont {Sramek}, \citenamefont {Fonteneau}, \citenamefont {Josse},\ and\ \citenamefont {Lucas}}]{SRAMEK1999189}%
  \BibitemOpen
  \bibfield  {author} {\bibinfo {author} {\bibfnamefont {R.}~\bibnamefont {Sramek}}, \bibinfo {author} {\bibfnamefont {G.}~\bibnamefont {Fonteneau}}, \bibinfo {author} {\bibfnamefont {E.}~\bibnamefont {Josse}}, \ and\ \bibinfo {author} {\bibfnamefont {J.}~\bibnamefont {Lucas}},\ }\bibfield  {title} {\enquote {\bibinfo {title} {Planar and channel waveguides in fluoride glasses},}\ }\href {\doibase https://doi.org/10.1016/S0022-3093(99)00521-9} {\bibfield  {journal} {\bibinfo  {journal} {Journal of Non-Crystalline Solids}\ }\textbf {\bibinfo {volume} {256-257}},\ \bibinfo {pages} {189--193} (\bibinfo {year} {1999})}\BibitemShut {NoStop}%
\bibitem [{\citenamefont {Miura}\ \emph {et~al.}(1999)\citenamefont {Miura}, \citenamefont {Qiu}, \citenamefont {Mitsuyu},\ and\ \citenamefont {Hirao}}]{MIURA1999212}%
  \BibitemOpen
  \bibfield  {author} {\bibinfo {author} {\bibfnamefont {K.}~\bibnamefont {Miura}}, \bibinfo {author} {\bibfnamefont {J.}~\bibnamefont {Qiu}}, \bibinfo {author} {\bibfnamefont {T.}~\bibnamefont {Mitsuyu}}, \ and\ \bibinfo {author} {\bibfnamefont {K.}~\bibnamefont {Hirao}},\ }\bibfield  {title} {\enquote {\bibinfo {title} {Preparation and optical properties of fluoride glass waveguides induced by laser pulses},}\ }\href {\doibase https://doi.org/10.1016/S0022-3093(99)00459-7} {\bibfield  {journal} {\bibinfo  {journal} {Journal of Non-Crystalline Solids}\ }\textbf {\bibinfo {volume} {256-257}},\ \bibinfo {pages} {212--219} (\bibinfo {year} {1999})}\BibitemShut {NoStop}%
\bibitem [{\citenamefont {Snyder}\ and\ \citenamefont {Love}(1983)}]{snyder1983optical}%
  \BibitemOpen
  \bibfield  {author} {\bibinfo {author} {\bibfnamefont {A.}~\bibnamefont {Snyder}}\ and\ \bibinfo {author} {\bibfnamefont {J.}~\bibnamefont {Love}},\ }\href {https://books.google.com.au/books?id=gIQB_hzB0SMC} {\emph {\bibinfo {title} {Optical Waveguide Theory}}},\ Science paperbacks\ (\bibinfo  {publisher} {Springer US},\ \bibinfo {year} {1983})\BibitemShut {NoStop}%
\bibitem [{\citenamefont {Gross}\ \emph {et~al.}(2015)\citenamefont {Gross}, \citenamefont {Jovanovic}, \citenamefont {Sharp}, \citenamefont {Ireland}, \citenamefont {Lawrence},\ and\ \citenamefont {Withford}}]{Gross2015}%
  \BibitemOpen
  \bibfield  {author} {\bibinfo {author} {\bibfnamefont {S.}~\bibnamefont {Gross}}, \bibinfo {author} {\bibfnamefont {N.}~\bibnamefont {Jovanovic}}, \bibinfo {author} {\bibfnamefont {A.}~\bibnamefont {Sharp}}, \bibinfo {author} {\bibfnamefont {M.}~\bibnamefont {Ireland}}, \bibinfo {author} {\bibfnamefont {J.}~\bibnamefont {Lawrence}}, \ and\ \bibinfo {author} {\bibfnamefont {M.~J.}\ \bibnamefont {Withford}},\ }\bibfield  {title} {\enquote {\bibinfo {title} {Low loss mid-infrared zblan waveguides for future astronomical applications},}\ }\href {\doibase 10.1364/OE.23.007946} {\bibfield  {journal} {\bibinfo  {journal} {Opt. Express}\ }\textbf {\bibinfo {volume} {23}},\ \bibinfo {pages} {7946--7956} (\bibinfo {year} {2015})}\BibitemShut {NoStop}%
\bibitem [{\citenamefont {B\'{e}rub\'{e}}, \citenamefont {Bernier},\ and\ \citenamefont {Vall\'{e}e}(2013)}]{Berube13}%
  \BibitemOpen
  \bibfield  {author} {\bibinfo {author} {\bibfnamefont {J.-P.}\ \bibnamefont {B\'{e}rub\'{e}}}, \bibinfo {author} {\bibfnamefont {M.}~\bibnamefont {Bernier}}, \ and\ \bibinfo {author} {\bibfnamefont {R.}~\bibnamefont {Vall\'{e}e}},\ }\bibfield  {title} {\enquote {\bibinfo {title} {Femtosecond laser-induced refractive index modifications in fluoride glass},}\ }\href {\doibase 10.1364/OME.3.000598} {\bibfield  {journal} {\bibinfo  {journal} {Opt. Mater. Express}\ }\textbf {\bibinfo {volume} {3}},\ \bibinfo {pages} {598--611} (\bibinfo {year} {2013})}\BibitemShut {NoStop}%
\bibitem [{\citenamefont {Ledemi}\ \emph {et~al.}(2014)\citenamefont {Ledemi}, \citenamefont {Bérubé}, \citenamefont {Vallée},\ and\ \citenamefont {Messaddeq}}]{Ledemi14}%
  \BibitemOpen
  \bibfield  {author} {\bibinfo {author} {\bibfnamefont {Y.}~\bibnamefont {Ledemi}}, \bibinfo {author} {\bibfnamefont {J.-P.}\ \bibnamefont {Bérubé}}, \bibinfo {author} {\bibfnamefont {R.}~\bibnamefont {Vallée}}, \ and\ \bibinfo {author} {\bibfnamefont {Y.}~\bibnamefont {Messaddeq}},\ }\bibfield  {title} {\enquote {\bibinfo {title} {Refractive index modification in fluoro-borate glasses containing wo3 induced by femtosecond laser},}\ }\href {\doibase https://doi.org/10.1016/j.jnoncrysol.2013.11.029} {\bibfield  {journal} {\bibinfo  {journal} {Journal of Non-Crystalline Solids}\ }\textbf {\bibinfo {volume} {385}},\ \bibinfo {pages} {153--159} (\bibinfo {year} {2014})}\BibitemShut {NoStop}%
\bibitem [{\citenamefont {Fernandez}\ \emph {et~al.}(2022)\citenamefont {Fernandez}, \citenamefont {Johnston}, \citenamefont {Gross}, \citenamefont {Cozic}, \citenamefont {Poulain}, \citenamefont {Mahmodi}, \citenamefont {Kabakova}, \citenamefont {Withford},\ and\ \citenamefont {Fuerbach}}]{Fernandez2022}%
  \BibitemOpen
  \bibfield  {author} {\bibinfo {author} {\bibfnamefont {T.~T.}\ \bibnamefont {Fernandez}}, \bibinfo {author} {\bibfnamefont {B.}~\bibnamefont {Johnston}}, \bibinfo {author} {\bibfnamefont {S.}~\bibnamefont {Gross}}, \bibinfo {author} {\bibfnamefont {S.}~\bibnamefont {Cozic}}, \bibinfo {author} {\bibfnamefont {M.}~\bibnamefont {Poulain}}, \bibinfo {author} {\bibfnamefont {H.}~\bibnamefont {Mahmodi}}, \bibinfo {author} {\bibfnamefont {I.}~\bibnamefont {Kabakova}}, \bibinfo {author} {\bibfnamefont {M.}~\bibnamefont {Withford}}, \ and\ \bibinfo {author} {\bibfnamefont {A.}~\bibnamefont {Fuerbach}},\ }\bibfield  {title} {\enquote {\bibinfo {title} {Ultrafast laser inscribed waveguides in tailored fluoride glasses: an enabling technology for mid-infrared integrated photonics devices},}\ }\href {\doibase 10.1038/s41598-022-18701-y} {\bibfield  {journal} {\bibinfo  {journal} {Scientific Reports}\ }\textbf {\bibinfo {volume} {12}},\ \bibinfo {pages} {14674} (\bibinfo {year} {2022})}\BibitemShut {NoStop}%
\bibitem [{\citenamefont {Kumar}, \citenamefont {Thyagarajan},\ and\ \citenamefont {Ghatak}(1983)}]{kumar1983analysis}%
  \BibitemOpen
  \bibfield  {author} {\bibinfo {author} {\bibfnamefont {A.}~\bibnamefont {Kumar}}, \bibinfo {author} {\bibfnamefont {K.}~\bibnamefont {Thyagarajan}}, \ and\ \bibinfo {author} {\bibfnamefont {A.~K.}\ \bibnamefont {Ghatak}},\ }\bibfield  {title} {\enquote {\bibinfo {title} {Analysis of rectangular-core dielectric waveguides: an accurate perturbation approach},}\ }\href@noop {} {\bibfield  {journal} {\bibinfo  {journal} {Optics Letters}\ }\textbf {\bibinfo {volume} {8}},\ \bibinfo {pages} {63--65} (\bibinfo {year} {1983})}\BibitemShut {NoStop}%
\bibitem [{\citenamefont {Gan}(1995)}]{gan1995optical}%
  \BibitemOpen
  \bibfield  {author} {\bibinfo {author} {\bibfnamefont {F.}~\bibnamefont {Gan}},\ }\bibfield  {title} {\enquote {\bibinfo {title} {Optical properties of fluoride glasses: a review},}\ }\href@noop {} {\bibfield  {journal} {\bibinfo  {journal} {Journal of non-crystalline solids}\ }\textbf {\bibinfo {volume} {184}},\ \bibinfo {pages} {9--20} (\bibinfo {year} {1995})}\BibitemShut {NoStop}%
\end{thebibliography}%

\end{document}